# Characteristics of binding sites of intergenic, intronic and exonic miRNAs with mRNAs oncogenes coding in-miRNAs

Olga A. Berillo[1], Assel S. Issabekova[1], Mireille Regnier[2], Anatoliy T. Ivashchenko[1a]

**Abstract:** Interaction of 784 intergenic, 686 intronic and 49 exonic miRNAs with mRNAs of 51 genes coding in-miRNAs were investigated. 43 genes are targets for ig-miRNAs, 44 for in-miRNAs and 17 for ex-miRNAs. These genes are affected by 92 ig-miRNAs, 128 in-miRNAs and 15 ex-miRNAs. Density of miRNA sites is higher in 5'UTR than it is in CDS and 3'UTR. Three types of miRNA interaction with mRNA were revealed. 5'-dominant canonical, 3'-compensatory and complementary types binding sites were revealed. Intronic miRNAs (in-miRNA) do not interact with mRNAs of host genes. Linkage between different mRNAs of genes encode in-miRNAs via other in-miRNAs are revealed. These data promote understanding of interaction mechanism of miRNA with mRNA genes participating in gastrointestinal cancer and development of microRNomics.

## Introduction

MicroRNAs (miRNA) are non-protein coding nucleotide sequences that have length about 22 nucleotides. They are evolutionary conservative in many organisms and accomplish important regulatory function (Ibanez-Ventoso et al., 2008). The binding miRNA to mRNA leads to specific splitting, deadenylation or translation repression (Pillai et al., 2007). Their number is considerably enlarged due to new methods of search and comparison of targets.

MiRNAs regulate translation of about 60% of all human non-protein-coding genes (Friedman et al., 2009). A mRNA can have some binding sites with one or with several miRNAs. Hence, the influence of miRNA on inhibition of mRNA translation becomes greater (Baek et al., 2008; Selbach et al., 2008; Grimson et al., 2007). Deregulation in miRNA expression is one main cause of cancer (Hamano et al., 2011), cardiovascular (Small et al., 2010) and other diseases (Jiang et al., 2010). Aberrant some miRNA expression have been well characterized in different oncology diseases, with implications for progression and prognosis (Cortez et al., 2012). Expression of some miRNAs changes in development esophageal cancer (Fang et al., 2012), gastric cancer (Lu et al., 2011), colorectal cancer (Hamfjord et al., 2012), breast cancer (Hafez et al., 2012; Hanna et al., 2012) and other types.

Recently, many programs for miRNA sites prediction have been developed (Sethupathy et al., 2006). A search for sites in the presence of complementary seed (part of 5'-end miRNA site) is carried out. The seed length depends on the program and ranges between 6 and 8 nt. (Friedman et al., 2009; Lewis et al., 2003). Site searching for only short seed leads to the prediction of considerable quantity of false sites. Some programs take advantage of seed conservation to assess site specificity. However there are many nonconservative seed-site of miRNAs already known (Baek et al., 2008). Some programs predict not perfectly complementary seed in the presence of 3'-compensatory site (Farh et al., 2005).

Most programs allow to search for sites only in 3'UTR (Maragkakis et al., 2011). Recently, it was shown that miRNA sites may also occur in ORF (Baek et al., 2008). Many researchers consider that miRNAs interact with mRNAs only in 3'UTR (Delay et al., 2011; Iorio et al., 2009; Isik et al., 2010; Kruger et al., 2006; Satoh et al., 2011), however publications which describe miRNA binding to mRNA in 5'UTR and CDS are known (Duursma et al., 2008; Elcheva et al., 2009; Kulkarni et al., 2011; Moretti et al., 2010; Tsai et al., 2009).

[1] Department of biology and biotechnology, al-Farabi Kazakh National University, Almaty 050038, Kazakhstan
[2] Inria, Centre de Recherche Saclay-Ile-de-France, Parc Orslay University, 91893, Orsay cedex, Paris, France
[a] – Corresponding author; e-mail: a_ivashchenko@mail.ru

The functionality of predicted sites according seed is checked experimentally, however other sites are not predicted by such method (Qin et al., 2010; Shirdel et al., 2011). In all cases, a miRNA site definition relies on the presence of some seed. Although some energy is calculated for a site, it is less important than the seed for the selection of a target mRNA. Program RNAhybrid relies on thermodynamics and allows to search for target sites subject to 5'-dominant initial, 5'-seed-dominant and 3'-compensatory types (Garcna et al., 2011). An accurate identification of miRNA:mRNA pairs should promote considerable development of diagnostic methods of various cancer diseases.

The present research aims to reveal binding miRNA sites with various parts of mRNA (5'UTR, CDS and 3'UTR) and define nucleotide interaction features of miRNA:mRNA complexes. This study is realized for mRNAs involved in cancer and should extend to other gene families.

**Materials and Methods**

Nucleotide sequences of miRNAs and their precursors were found from miRBase (http://www.mirbase.org). mRNA sequences of all genes (*Homo sapiens*, Genome build 37.2.) were searched in obtained from GenBank (http://www.ncbi.nlm.nih.gov). Then, miRNAFinder 2.2 (http://sites.google.com/site/malaheenee/software/) was used to find miRNA origins (intergenic, exonic or intronic). A literature review of genes coding intronic miRNAs led to 51 genes (Table S. 1) that encode proteins participating in gastrointestinal and breast cancer. These information was collected according to articles from US National Library of Medicine National Institutes of Health PubMed (http://www.ncbi.nlm.nih.gov/pubmed) (S. references).

Software RNAHybrid 2.1. was run for all pairs (miRNA:mRNA) and provided postions of potential binding sites miRNAs in mRNAs, their energies $\Delta G$ and scheme of their interaction. Our script E-RNAhybrid (http://sites.google.com/site/malaheenee/software/) computes the ratio $\Delta G/\Delta G_m$, p-value, equalizing coefficient and type corresponding regions (5'UTR, CDS or 3'UTR), where miRNA site disposes. To achieve this, a quantitative criterion is defined. A ratio $\Delta G/\Delta G_m$ value (%), where $\Delta G_m$ equals binding energy for miRNA with perfectly complementary nucleotide sequence, is computed. Number of binding sites in 5'UTR, CDS and 3'UTR was computed as the number of sites divided to nucleotide length of this region and multiplied by $10^3$ (s/l), that is calculation per 1000 nucleotides. Reliability degree (p-value) is estimated, that relies on $\Delta G$ and its standard deviation. Threshold significance is set to $p<0.0005$.

Energy of miRNA binding site with mRNA depends on number of complementary nucleotide pairs. The miRNA length is an important parameter in selection of binding sites. The miRNA length varies in range 16-27 n. If the length miRNA is shorter, the probability of receiving false positive results is higher, so the equalizing coefficient was entered. It was revealed according to distribution of miRNA length that the majority miRNAs have 21 n., which was it is taken as basic value. Values of equalizing coefficient was calculated in the searching of all studied miRNA sites on casual nucleotide sequences, that have average length of mRNAs. If the miRNA length is more than 21 n., value of threshold significance (p) divided into equalizing coefficient, thus reliability becomes higher. In case, the miRNA length is less than 21 n., value of threshold significance (p) multiplied by the corresponding coefficient, thus reliability becomes less.

**Results**

784 intergenic, 686 intronic and 49 exonic miRNAs were found by program miRNAFinder 2.2. 471 genes encode 686 intronic pre-miRNAs. Each gene encode one or more in-miRNAs. Among them only 51 encode proteins participating in development of gastrointestinal and breast cancer according to literature review of all these genes. Characteristics 51 genes are represented in Table S. 1. Five genes (*ATF2, BID, DMD, EVL, FOXP1*) encode intronic pre-miRNAs in their 3'UTRs and another five genes (*BBC3, EPHB2, NR2F2, PTPRJ, SPATA13*) encode intronic pre-miRNAs in their 5'UTRs. Some genes encode one or more in-miRNA. For example, *AATK* gene encodes pre-miR-338, pre-miR-657 and pre-miR-1250; *HDAC4* gene encodes pre-miR-2467, pre-miR-4440 and pre-miR-4441.

Interaction of all intergenic, intronic and exonic miRNAs with mRNAs of 51 genes were investigated by programs RNAHybrid 2.1 and E-RNAhybrid. We show a linkage between different mRNA of genes encode in-miRNAs via other in-miRNAs (Table S. 3). Binding sites to miRNAs according to threshold significance ($p < 0.0005$) with equalizing coefficient (Table S. 4 – 8). Similar results can be found on three types of miRNA. We show that binding may involve any mRNA region (5'UTR, CDS and 3'UTR). The density of sites and their distribution on these mRNA regions were revealed. We consider an example, *HDAC4* mRNA that has several types of binding sites completely. Complementary type of miRNA binding sites was found after analyze of our results, by this mean we improve a classification of miRNA binding sites according to their localization.

*Interaction of intergenic miRNA with mRNA genes coding in-miRNA*

Binding sites of 784 ig-miRNAs with these mRNAs were studied. 43 mRNAs were found to be targets for ig-miRNAs and 8 mRNAs (*ABCF1, BIRC6, DCC, EBF3, FBXW7, IGF2, SLIT2, TNFAIP6*) have no binding sites to ig-miRNAs ($p < 0.0005$). Only 92 out of 784 ig-miRNAs are involved (Table S. 4, 5). The number and density of binding sites of ig-miRNAs to mRNAs depend on the mRNAs.

Shares of binding sites for ig-miRNAs are 15.8%, 57.7% and 26.5% in 5'UTR, CDS, 3'UTR of 49 mRNAs, respectively (Figure 1, A). However, the average number of binding sites in 5'UTR, CDS and 3'UTR mRNA is 2.54 s/l, 0.99 s/l and 0.90 s/l, respectively for these genes (Figure 1, B). Table S. 4 represents the data for the 35 genes with a small number (one to six) of ig-miRNA sites. These ig-miRNAs have only one site in mRNA-targets. Data for 8 remaining genes, that have seven or more binding sites to ig-miRNAs are given in Table S. 5. Some mRNAs have more than one interaction site with one ig-miRNA. For example, miR-4472 has four sites in *LRP1* mRNA and two sites in *NOTCH1* and *AATK* mRNAs. The site number ranges from 0.19 s/l (*ABCA6* mRNA) to 5.47 s/l (*BBC3* mRNA) with an average value of 0.94 s/l. (Table S. 9) The highest numbers are found for *AATK, BBC3, HDAC4, HUWE1, IGF1R, LRP1, NOTCH1* and *SLIT3* mRNAs, with 13, 10, 13, 7, 8, 12, 8 and 10 binding sites, respectively.

Ig-miRNA sites with the best characteristics are represented in Tables 1 – 3, there are results with the ratio $\Delta G/\Delta G_m$ 80% and more (with equalizing coefficient). Table 1 shows characteristics of some ig-miRNA sites in 5'UTR. The ratio $\Delta G/\Delta G_m$ of these miRNA sites changes in diapason 80-92%. Translation some mRNAs can be regaled by different miRNA. For example, *NR2F2* gene have four binding sites; *EPCAM*, *HNF4A* and *SLIT3* genes have three sites and *EGFL7, HDAC4, LRP1* genes has two sites in 5'UTR. Table 2 shows ig-miRNA sites in CDS with ratio $\Delta G/\Delta G_m$ 80-99%. Many of mRNAs have some miRNA sites. MiR-4472 has binding sites in CDS of 7 genes (*AATK, HDAC4, LRP1, MAP2K4, NOTCH1, PTPRJ, SLIT3*), where *LRP1* gene has three sites and *NOTCH1* gene has two sites. MiR-4456, miR-4455, miR-4711-3p have bindings sites in CDS six, five, four mRNAs according. These miRNAs are important as they have binding sites with many mRNAs of studied genes. Table 3 shows ig-miRNA sites in 3'UTR with ratio $\Delta G/\Delta G_m$ 81-84%. Characteristics of site location same as in 5'UTR and CDS. *BRE, IGF1R* and *MRE11A* genes are targets for miR-4456; *BBC3* and *NOTCH1* genes are targets for miR-1587. These data correspond to reliability $p < 0.0004$. There are miRNAs, their mRNA-targets and position of each binding sites. Majority of these miRNAs were revealed recently (with miRNA index number 1000 and more) and low-studied.

*Interaction of intronic miRNAs with mRNAs genes coding in-miRNA*

In-miRNAs are coded in introns of pre-mRNA, introns of 5'UTR and introns of 3'UTR. Binding sites of 686 in-miRNAs with 51 mRNAs were studied (Tables S. 6, 7). 44 mRNAs were found to be potential targets according to our threshold significance ($p < 0.0005$). Only 128 out of the 686 in-miRNAs are involved in the interactions. Seven mRNAs (*ABCF1, ATF2, EPHB2, HNF4A, MRE11A, SDCCAG8, TNFAIP6* genes) have no binding sites to in-miRNAs. Only miR-1268b has two binding sites to *BBC3* mRNA.

The number of connection with miRNA are significantly different for studied mRNAs. 12 mRNAs have sites only with one miRNA (Table S. 6). mRNAs with seven and more binding sites are depicted in Table S. 7. The average number of all in-miRNA sites is 2.9. mRNAs of *AATK, AKT, HDAC4, IGF2* and *LRP1* genes have 10, 11, 13, 7 and 14 binding sites, respectively. Some mRNAs have more than one binding site with a given in-miRNAs. For example, miR-1268b and miR-4296 have two binding sites with *HDAC4* mRNA. miR-1273f and miR-574-5p have two and five binding sites with *IGF2* mRNA.

Density of in-miRNA binding sites in 44 mRNAs are represented in Table S. 10. The miRNA site density ranges from 0.17 s/l for *AXTXR1* mRNA to 2.74 s/l for *BBC3* mRNA.

5'UTR, CDS and 3'UTR of mRNA of studied genes have different binding ability to miRNA. in-miRNAs shares are 16.1%, 46.4% and 37.5% sites in 5'UTR, CDS, 3'UTR of 44 mRNA, respectively (Figure 1, B). However, the average binding sites densities of 5'UTR, CDS and 3'UTR computed over all mRNAs are 1.82 s/l, 0.63 s/l and 0.72 s/l, respectively (Figure 1, B). Average density of binding sites for 5'UTR is 2.9 times more than it is in CDS and 2.6 times more than it is in 3'UTR. These results indicate that miRNAs can bind to 5'UTR, CDS and 3'UTR.

In-miRNA sites with the best characteristics are represented in Table 4, there are results with the ratio $\Delta G/\Delta G_m$ 80% ($p < 0.0004$) and more (with equalizing coefficient). Location of miRNA site depend on mRNA gene. For example, *IGF1R, LRP1* and *IGF2* genes are targets for miR-1273f and its site locates in 3'UTR, CDS and 5'UTR corresponding. Characteristics of 81 in-miRNA sites for 33 mRNAs of target genes encoding other intronic miRNAs in Table S. 3. There are characteristics of miRNA binding sites and their origin (gene, where encodes pre-miRNA). These in-miRNAs have different origin their pre-miRNAs. Among all in-miRNA, four pre-miRNA encode in introns of 3'UTR and 20 pre-miRNAs in introns of 5'UTR. *LRP1* gene has six miRNA sites in different domains of its mRNA: CDS (miR-1273f, miR-4295, miR-4296, miR-500b), 5'UTR (miR-4274) and 3'UTR (miR-4297). *HUWE1* gene has sites (miR-4297, miR-548an, miR-598) only in CDS and *LFNG* gene has sites (miR-1224-3p, miR-500b) only in 3'UTR. Some mRNAs can be target to in-miRNAs, that encode in different pre-mRNA genes. For example, *IGF1R* gene is target for miR-1268B (*CCDC40*), miR-1273F (*SCP2*), miR-3173 (*DICER1*), miR-361 (*CHM*), miR-4292 (*C9orf86*), miR-4297 (*EBF3*), where gene origin corresponding pre-miRNA is shown in brackets. These data show a linkage between different genes encode intronic miRNAs, where each mRNA is target for some miRNAs and encodes other miRNAs. By this means, such genes can affect with each other via in-miRNAs.

*Interaction of exonic miRNAs with mRNA genes coding in-miRNA*

49 ex-miRNAs are coded in exons of pre-mRNAs, 5'UTRs and 3'UTRs. Binding sites of ex-miRNAs with 51 mRNAs were studied (Table S. 8). Only 15 out of 49 ex-miRNAs interact with no more than 17 target mRNAs. Shares of binding sites for ex-miRNAs are 14.3%, 57.1% and 28.6% sites in 5'UTR, CDS, 3'UTR of 17 mRNA, respectively (Figure 1, A). The number of binding sites of ex-miRNAs with mRNAs of target genes are different. For example, *EPHB2* mRNA has three sites one in 5'UTR and two in CDS), and *DMD* and *FOXP1* mRNAs have two sites. There are no cases, where mRNA have two and more binding sites.

5'UTR, CDS and 3'UTR of these genes mRNAs have different binding ability to ex-miRNA. The average binding sites density of miRNA in 5'UTR, CDS and 3'UTR equals 0.64 s/l, 0.29 s/l and 0.18 s/l, respectively (Figure 1, B). Average binding sites density for 5'UTR is 2.2 times more than it is in CDS and 3.6 times more than it is in 3'UTR. The sites density ranges from 0.06 s/l for *BIRC6* mRNA to 0.76 s/l for *BIRC7* mRNA with an average value 0.29 s/l (Table S. 11). in-miRNA sites with the best characteristics are represented in table 5, there are results with the ratio $\Delta G/\Delta G_m$ 80% ($p<0.0004$) and more (with coefficient equalizing). Despite the small number of target genes, these data have proved that ex-miRNA can bind to 5'UTR, CDS and 3'UTR.

*HDAC4 mRNA has three types of miRNA binding sites.*

As is known, it exist three types of binding sites: 5'-dominant canonical, 5'-dominant seed only and 3'-compensatory (Betel et al., 2010). Canonical site has good (or perfect) complementarity at both the 5'- and 3'-ends of the miRNA with a specific bulge in the middle. Dominant seed site has perfect seed 5'-complementarity to the miRNA but poor 3'-complementarity. Compensatory site has a mismatch (wobble) in the 5'-seed region but compensate through excellent complementarity at the 3'-end.

5'-dominant canonical and 3'-compensatory types of binding sites were found in our data. There is not 5'-dominant seed only type of binding sites, because such sites have low energy of binding site and are not reliable according to our criteria. In addition, complementary type of miRNA binding sites was found after analyze of our results. It has perfect complementarity beginning with second nucleotide and completing the last but one nucleotide of mRNA in binding site. There is not a bugle in the middle like in canonical site (Table 6). Such binding sites have high energy of binding site and are reliable according to our criteria.

*HDAC4* mRNA has three types of miRNA binding sites. An example, 5'-dominant canonical sites are in-miR-4296 and ig-miR-3195; 3'-compensatory sites are ig-miR-4311 and in-miR-1914*; complementary sites are ig-miR-4478 and in-miR-1289 (Table 6). The primary contribution to energy can include all parts of miRNA site but not only 5'seed.

*HDAC4* mRNA is a target for several intergenic and intronic miRNAs. In both cases, sites may be found in any region. *HDAC4* mRNA has seven, nine and seven sites for intergenic miRNAs in 5'UTR, CDS and 3'UTR, respectively (Table S. 4). It has eight, two and five sites for intronic miRNAs in 5'UTR, CDS and 3'UTR, respectively (Table S. 7). A specific attention is paid to two miRNAs (miR-1268b and miR-4296) that have two binding sites with *HDAC4* mRNA. In both cases, two homologous sites can be found in 5'UTR.

Two sites of miR-4787-5p are located in 5'UTR of HDAC4 mRNA and they are highly homological. miR-1587 and miR-4507 are placed in the same site *HDAC4* mRNA (Table S. 5). These examples indicate a highly specific interaction between miRNAs and mRNAs. Found sites have high hybridization energy due to significant amount of complementary nucleotides.

**Discussion**

According to (Friedman et al., 2009), expression may be regulated by miRNAs for more than a half of human genes. The number of known human miRNAs constantly increases, as well as the number of potential target genes for them. The establishment of target genes depends on efficiency of their prediction by computer methods. Indeed, experimental verification of predicted target genes depends on accuracy of prediction of miRNA sites and their characteristics considered in corresponding programs. Besides, the essential role of prediction of binding sites and their validation have limitation depending on quality of prediction (Grimson et al., 2007).

In earlier publications, several tens or hundreds miRNAs were investigated by researchers. In our search work, interaction sites of 1519 miRNAs with 51 mRNA-targets are presented. Total number of miRNA sites found in all target genes is 439. Some miRNAs have several binding sites with the same mRNA. Average number of binding sites is 6 miRNAs on mRNA. MiRNA sites with high ratio $\Delta G/\Delta G_m$ value were selected with $p < 0.0005$. In-miRNAs and ig-miRNAs have similar properties. miRNAs with length 22 n. have distinct prevalence concerning others which have bigger or smaller length. A great number of in-miRNAs and ig-miRNAs have a high GC-content (50-55%). Most genes under study appear to have binding sites with miRNA. Therefore, it is possible that the expression of a significant part of human genes is under a regulation control by miRNAs.

Ig-miR-4456, ig-miR-4455, ig-miR-4492, ig-miR-302f, ig-miR-3195 have bindings sites in CDS eight, six, five and five mRNAs according. In-miR-1268b, in-miR-4753-3p, in-miR-574-5p have bindings sites in CDS six, four and four mRNAs according. The expression of such miRNAs can repress translation of many genes, that participate in development cancer.

The shares of all binding sites 15.8%, 53.1% and 31.1% in 5'UTR, CDS, 3'UTR of 51 mRNA respectively were found (Figure 1, A). Shares of in-miRNA and ig-miRNA that have binding sites in 5'UTR, CDS and 3'UTR are similar. However the average number of in-miRNA sites towards single mRNA is 19% more than it is to ig-miRNA.

Thus, the density of binding sites in 5'UTR is 2.6 times more than it is in CDS and 2.8 times more than it is in 3'UTR. Such data indicate that ig-miRNAs can interact with 5'UTR and CDS, and not only with 3'UTR. Some miRNAs have a high site density in 5'UTR.

A significant part of studied miRNAs interact with mRNA of only one gene. Action on such miRNAs would allow for a selective modification of associated target genes expression. MRNA of some genes is targets for ig-miRNAs and in-miRNAs.

### Conclusions

Computational prediction of miRNA binding sites is important stage in investigation biological function miRNA. These data promote understanding features of gene regulation on post-transcription level via microRNAs. The linkage between different genes encode in-miRNAs and are targets for in-miRNAs were revealed. These genes participate in different cellular processes and revealing linkages between such genes participating in gastrointestinal cancer.

It has been found that the percentage of ig-miRNA, ex-miRNA, in-miRNA sites are approximately identical in 5'UTR, CDS, 3'UTR. Approximately 2/3 of these sites bind to 5'UTR or CDS regions. This suggests that sites involved in a translation regulation by miRNAs are located in 5'UTR and CDS regions, and not only in 3'UTR.

On these data, around 45% of miRNAs are intronic and their synthesis directly depends on transcription of corresponding host genes. In-miRNAs have no strong binding to mRNAs of 51 studied genes. Therefore in-miRNA do not inhibit the expression of host genes.

Three types of schemes of interaction between mRNA and miRNA were revealed. There are 5'-dominant canonical sites, 3'-compensatory sites and complementary sites. The primary contribution to energy can include all parts of miRNA site but not only 5'seed. Hence all parts of miRNA sites can bring contribution to the total energy of site ($\Delta G/\Delta G_m$).

### Conflict of interest statement
None declared.

### Acknowledgements
This work was supported by a grant «Establishment correlation between microRNA expression, their target genes and human breast cancer development. Developing method of early disease diagnostics» (2012-2014) from the Ministry of Education and Science, Republic of Kazakhstan.

We thank Vladimir Khailenko for creating programs [miRNAFinder 2.2](miRNAFinder 2.2) and E-RNAhybrid.

### Author Disclosure Statement
The author declares that no conflicting financial interests exist.

A

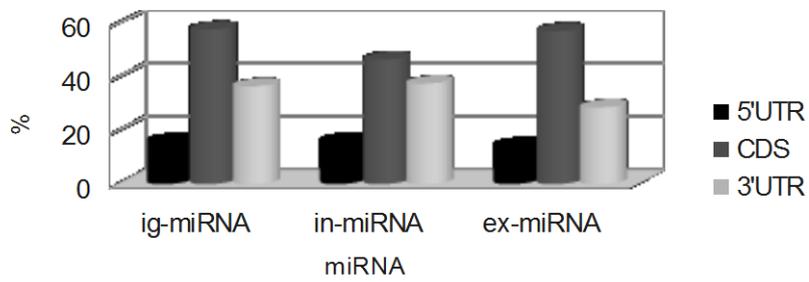

B

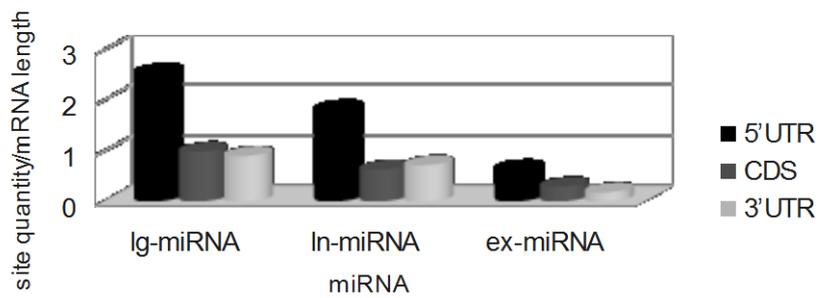

*Note:* These binding sites have ratio ΔG/ΔG$_m$ equal 70–99%.
A – Shares of miRNAs binding sites with different parts of 51 mRNAs.
B – Average number of miRNA binding sites with different parts of 51 mRNAs.

**Fig. 1.** Characteristics of miRNA binding sites with different parts of all mRNAs

Table 1. Characteristics of sites binding ig-miRNA with 5'UTR mRNA

| Gene | miRNA | Position | Gene | miRNA | Position | Gene | miRNA | Position |
|---|---|---|---|---|---|---|---|---|
| AATK | miR-4417 | 45 | HDAC4 | miR-1268 | 80 | NR2F2 | miR-4253 | 537 |
| ATF2 | miR-4319 | 63 | HNF4A | miR-302f | 71 | | miR-4787-5p | 1152 |
| BRE | miR-4455 | 39 | | miR-1279 | 1 | PRKG1 | miR-1268 | 142 |
| DCC | miR-568 | 524 | | miR-4307 | 76 | PTK2 | miR-3676 | 22 |
| DTL | miR-4309 | 44 | HUWE1 | miR-4266 | 134 | PTPRJ | miR-3195 | 288 |
| EGFL7 | miR-4289 | 4 | LRP1 | miR-4307 | 188 | SDCCAG8 | miR-4309 | 20 |
| | miR-1204 | 301 | | miR-4472 | 99 | SLIT3 | miR-4507 | 61 |
| EPCAM | miR-4456 | 112 | MCM7 | miR-4289 | 453 | | miR-4466 | 253 |
| | miR-4492 | 137 | MTUS1 | miR-3195 | 141 | | miR-4481 | 314 |
| | miR-4508 | 140 | NR2F2 | miR-4443 | 556 | SPATA13 | miR-4279 | 12 |
| HDAC4 | miR-3195 | 266 | | miR-1538 | 289 | | | |

*Note:* These binding sites have ratio $\Delta G/\Delta G_m$ equal 80-92%

Table 2. Characteristics of sites binding ig-miRNA with CDS mRNA

| Gene | miRNA | Position | Gene | miRNA | Position | Gene | miRNA | Position |
|---|---|---|---|---|---|---|---|---|
| AATK | miR-3195 | 1352 | EPHB2 | miR-4316 | 1293 | LRP1 | miR-4455 | 12015 |
|  | miR-4472 | 1684 |  | miR-4466 | 1334 |  | miR-4266 | 5093 |
|  | miR-4711-3p | 2715 | ERBB4 | miR-4443 | 1940 |  | miR-132 | 4503 |
|  | miR-4492 | 3345 | EVL | miR-4711-3p | 802 |  | miR-4472 | 13015 |
|  | miR-1538 | 1339 |  | miR-3180 | 816 |  | miR-4456 | 2930 |
|  | miR-4265 | 3143 | FOXP1 | miR-4736 | 1146 |  | miR-1261 | 7704 |
| AKT2 | miR-4492 | 1370 |  | miR-4266 | 1126 |  | miR-4253 | 10550 |
| ANTXR1 | miR-4711-3p | 1772 |  | miR-320d | 2225 | MAP2K4 | miR-4472 | 120 |
|  | miR-3141 | 2010 |  | miR-4264 | 2506 | MAP7D2 | miR-4417 | 1395 |
| BBC3 | miR-4483 | 744 |  | miR-1279 | 2111 | MCM7 | miR-4483 | 1975 |
|  | miR-4710 | 796 |  | miR-4327 | 2392 | MRE11A | miR-520e | 693 |
|  | miR-3665 | 410 | GIPR | miR-659 | 1441 |  | miR-302f | 852 |
|  | miR-4466 | 319 | HDAC4 | miR-4483 | 2802 | NOTCH1 | miR-4472 | 4308 |
|  | miR-4278 | 884 |  | miR-4481 | 1868 |  | miR-4472 | 7474 |
| BCAS1 | miR-4307 | 871 |  | miR-302e | 2082 |  | miR-4455 | 6381 |
| BID | miR-596 | 364 |  | miR-4472 | 2159 |  | miR-4455 | 6997 |
| BIRC7 | miR-4456 | 372 |  | miR-1205 | 976 |  | miR-516a-3p | 6964 |
| BRE | miR-3201 | 878 |  | miR-4746-3p | 982 |  | miR-516b* | 6964 |
| CCAR1 | miR-548ak | 3272 | HNF4A | miR-1204 | 278 |  | miR-4736 | 75 |
|  | miR-4463 | 841 |  | miR-4456 | 465 |  | miR-4283 | 1245 |
| CDH13 | miR-4455 | 2032 | HUWE1 | miR-320c | 3048 | PRKG1 | miR-599 | 2933 |
| DCC | miR-4711-3p | 3297 |  | miR-4307 | 3373 | PTPRJ | miR-4472 | 2750 |
|  | miR-1207-3p | 3190 |  | miR-4264 | 6886 | SLIT3 | miR-302e | 3310 |
|  | miR-4318 | 1684 |  | miR-1279 | 10486 |  | miR-4472 | 3745 |
|  | miR-4325 | 2917 |  | miR-320d | 3050 |  | miR-302f | 3310 |
| DMD | miR-4282 | 2958 |  | miR-4443 | 4959 |  | miR-4513 | 499 |
|  | miR-4493 | 5132 |  | miR-4792 | 10624 |  | miR-4508 | 1201 |
|  | miR-4520a-5p | 1878 | IGF1R | miR-4456 | 462 |  | miR-4481 | 4872 |
|  | miR-4520b-5p | 1878 |  | miR-4282 | 528 |  | miR-4492 | 1201 |
|  | miR-4650-5p | 5135 |  | miR-1268 | 1535 | SPATA13 | miR-1261 | 1854 |
| DNMT3A | miR-302f | 3075 | LFNG | miR-4264 | 1154 |  | miR-4266 | 2070 |
|  | miR-3676 | 1221 |  | miR-1274b | 175 |  | miR-4456 | 2864 |
|  | miR-665 | 1570 |  | miR-3195 | 178 | TNKS | miR-4455 | 3891 |
|  | miR-4456 | 2223 | LRP1 | miR-4472 | 11274 |  | miR-4531 | 647 |
| EIF4H | miR-645 | 705 |  | miR-4472 | 2781 |  | miR-4465 | 1769 |
| EPHB2 | miR-4253 | 1087 |  | miR-4283 | 12932 |  |  |  |

*Note:* These binding sites have ratio $\Delta G/\Delta G_m$ equal 80-99%

Table 3. Characteristics of sites binding ig-miRNA with 3'UTR mRNA

| Gene | miRNA | Position | Gene | miRNA | Position | Gene | miRNA | Position |
|---|---|---|---|---|---|---|---|---|
| AATK | miR-4472 | 4514 | EPHB2 | miR-4455 | 4792 | IGF1R | miR-4455 | 8927 |
| AKT2 | miR-4316 | 2472 | ERBB4 | miR-4279 | 10500 | | miR-769-3p | 7824 |
| | miR-4418 | 3951 | | miR-568 | 11012 | | miR-4455 | 10709 |
| | miR-4264 | 2336 | | miR-3123 | 4973 | LFNG | miR-466 | 1269 |
| BBC3 | miR-4507 | 1000 | | miR-302f | 5163 | | miR-4318 | 1899 |
| | miR-4505 | 1000 | | miR-513a-5p | 4123 | LRP1 | miR-4328 | 14372 |
| | miR-4497 | 965 | FBXW7 | miR-3674 | 3468 | MAP2K4 | miR-1827 | 3050 |
| | miR-3676 | 1616 | FOXP1 | miR-466 | 5945 | MCM7 | miR-4466 | 2782 |
| | miR-1587 | 1000 | | miR-4266 | 3217 | MRE11A | miR-4456 | 3832 |
| | miR-4279 | 1636 | HDAC4 | miR-4478 | 8388 | MTUS1 | miR-513a-5p | 4652 |
| BID | miR-543 | 2428 | | miR-4529-5p | 6541 | | miR-4443 | 6051 |
| BRE | miR-4456 | 1575 | | miR-4482 | 7344 | | miR-3168 | 4674 |
| CDH13 | miR-297 | 3603 | | miR-4710 | 8162 | | miR-513b | 4652 |
| DCC | miR-302f | 6493 | | miR-4311 | 7024 | NOTCH1 | miR-1587 | 8133 |
| EGFL7 | miR-3130-5p | 1454 | | miR-4328 | 4662 | | miR-1275 | 9070 |
| EIF4H | miR-4309 | 1364 | HNF4A | miR-3934 | 1350 | SPATA13 | miR-876-3p | 6126 |
| | miR-197 | 1627 | IGF1R | miR-4456 | 7101 | | | |

*Note:* These binding sites have ratio $\Delta G/\Delta G_m$ equal 80-94%

Table 4. Characteristics of sites binding in-miRNA with 5'UTR, CDS, 3'UTR mRNA

| Gene | miRNA | Position | Gene | miRNA | Position | Gene | miRNA | Position |
|---|---|---|---|---|---|---|---|---|
| **5'UTR** | | | DNMT3A | miR-593 | 674 | BBC3 | miR-3156-3p | 1667 |
| BIRC7 | miR-4292 | 60 | ERBB4 | miR-3182 | 3863 | CDH13 | miR-574-5p | 3603 |
| EPCAM | miR-4753-3p | 7 | | miR-4797-3p | 3376 | EBF3 | miR-500b | 2252 |
| | miR-4317 | 45 | EVL | miR-1322 | 522 | | miR-1976 | 2260 |
| GIPR | miR-26a-1* | 25 | | miR-4281 | 1001 | EGFL7 | miR-3130-5p | 1454 |
| HDAC4 | miR-1914* | 540 | GIPR | let-7g | 140 | EIF4H | miR-4263 | 1611 |
| | miR-1268b | 78 | | miR-1238 | 182 | ERBB4 | miR-483-3p | 10500 |
| | miR-4296 | 53 | LRP1 | miR-1273f | 3301 | | miR-877* | 10504 |
| | miR-4296 | 159 | | miR-4295 | 6360 | HDAC4 | miR-1289 | 6515 |
| IGF2 | miR-1273f | 81 | | miR-500b | 1824 | | miR-4326 | 4606 |
| LRP1 | miR-4274 | 428 | NOTCH1 | miR-3196 | 688 | IGF1R | miR-4292 | 6166 |
| PRKG1 | miR-1268b | 140 | | miR-1271 | 1240 | | miR-3173-5p | 5679 |
| PTPRJ | miR-1238 | 208 | PTPRJ | miR-342-5p | 459 | | miR-361-3p | 7089 |
| **CDS** | | | SLIT2 | miR-4446-3p | 4223 | | miR-1273f | 7624 |
| AATK | miR-4651 | 4029 | SLIT3 | miR-500b | 426 | IGF2 | miR-1273f | 1902 |
| ABCA6 | miR-3941 | 3053 | SPATA13 | miR-652 | 482 | | miR-4263 | 3055 |
| BBC3 | miR-4655-3p | 598 | **3'UTR** | | | | miR-3972 | 4533 |
| BID | miR-4285 | 432 | AATK | miR-648 | 4498 | | miR-4296 | 4307 |
| BIRC7 | miR-4257 | 745 | AKT2 | miR-3196 | 1865 | LFNG | miR-1224-3p | 2007 |
| BRE | miR-4259 | 315 | | miR-4281 | 2711 | | | |
| DMD | miR-4753-3p | 11190 | ANTXR1 | miR-32* | 5217 | | | |

*Note:* These binding sites have ratio $\Delta G/\Delta G_m$ equal 80-91%

Table 5. Characteristics of sites binding ex-miRNA with 5'UTR, CDS, 3'UTR mRNA

| Gene | miRNA | Position | Gene | miRNA | Position | Gene | miRNA | Position |
|---|---|---|---|---|---|---|---|---|
| | **CDS** | | DMD | miR-1306 | 2819 | | **3'UTR** | |
| BIRC6 | miR-4315 | 11064 | GIPR | miR-671-5p | 1236 | DNMT3A | miR-4775 | 3420 |
| BIRC7 | miR-1825 | 373 | HUWE1 | miR-3652 | 6967 | MTUS1 | miR-1306 | 5360 |
| DMD | miR-4315 | 6086 | | | | | | |

*Note:* These binding sites have ratio $\Delta G/\Delta G_m$ equal 81-84%

Table 6. Schematic representation of types of bindings sites miRNAs with mRNA *HDAC4*

```
5'UTR,53     ΔG=-33,5    ΔG/ΔGm=84    5'-d/c      5'UTR,266    ΔG=-38,8    ΔG/ΔGm=84    5'-d/c

mRNA         5' C      C           C 3'          mRNA         5' C      C           C 3'
                GAGCC  GAGCCCGCG                                 AGCCC  GGCCCGGCGC
                CUCGG  CUCGGGUGU                                 UUGGG  CCGGGCCGCG
in-miR-4296  3' A      A           A 5'          ig-miR-3195  3'        C           C 5'

3'UTR,7024   ΔG=-28,8    ΔG/ΔGm=80    3'-compn.   5'UTR,540    ΔG=-46,8    ΔG/ΔGm=81    3'-compn.

mRNA         5' A                C       3'      mRNA         5'G                    G G   C 3'
                CACACUCGGCUCUU                                  UCUCCCGGUGCGGGGCCC C CC
                GUGUGAGUCGAGAG                                  GGAGGGUCACGCCCUGGG G GG
ig-miR-4311  3'                 AAAG 5'          in-miR-1914* 3'                      A     5'

3'UTR,8388   ΔG=-34,6    ΔG/ΔGm=87    compl.      3'UTR,6515   ΔG=-36,7    ΔG/ΔGm=80    compl.

mRNA         5' G                    A 3'        mRNA         5' U                          A 3'
                UUCUUAGCUCGGCCUC                                 GAGUGCAGAUUCUUGGAUUC
                AGGAGUCGAGUCGGAG                                 UUUACGUCUAAGGACCUGAG
ig-miR-4478  3' G                    5'          in-miR-1289  3' U                         GU 5'
```

*Note*. Site localization of 5'-binding sites of nucleotide interaction miRNAs with mRNAs are shown. Interaction energy (ΔG) is measured in kcal/mol. The ΔG/ΔG$_m$ value is measured in percents. Types of bindings sites: 5'-d/c – 5'-dominant canonical, 3'-comp. – 3'-compensatory, compl. – complementary site.